\documentclass[aps, prb, 12pt, letterpaper, superscriptaddress]{revtex4-2}

\usepackage[utf8]{inputenc}
\usepackage[english]{babel}
\usepackage{hyperref}
\usepackage{amsmath}
\usepackage{amssymb}
\usepackage{graphicx}
\usepackage{color}
\usepackage{bm}
\usepackage{float}

\begin{document}

\title{Magnetic field driven dynamics in twisted bilayer artificial spin ice at superlattice angles}

\author{Rehana Begum Popy}
\affiliation{Department of Physics and Astronomy, University of Manitoba, Winnipeg, Manitoba, R3T 2N2, Canada}
\author{Julia Frank}
\affiliation{Department of Physics and Astronomy, University of Manitoba, Winnipeg, Manitoba, R3T 2N2, Canada}
\author{Robert L. Stamps}
\affiliation{Department of Physics and Astronomy, University of Manitoba, Winnipeg, Manitoba, R3T 2N2, Canada}

\date{\today}

\begin{abstract}
Geometrical designs of interacting nanomagnets have been studied extensively in the form of two dimensional arrays called artificial spin ice. These systems are usually designed to create geometrical frustration and are of interest for the unusual and often surprising phenomena that can emerge. Advanced lithographic and element growth techniques have enabled the realization of complex designs that can involve elements arranged in three dimensions. Using numerical simulations employing the dumbbell approximation, we examine possible magnetic behaviours for bilayer artificial spin ice (BASI)  in which the individual layers are rotated with respect to one another. The goal is to understand how magnetization dynamics are affected by long-range dipolar coupling that can be modified by varying the layer separation and layer alignment through rotation. We consider bilayers where the layers are both either square or pinwheel arrangements of islands. Magnetic reversal processes are studied and discussed in terms of domain and domain wall configurations of the magnetic islands. Unusual magnetic ordering is predicted for special angles which define lateral spin superlattices for the bilayer systems. 
\end{abstract}

\maketitle


 
\section{Introduction}
 
Artificial spin ice (ASI) are lithographically designed systems consisting of interacting nanomagnets arranged to create effects associated with geometrical frustration. A great advantage of studying experimentally these systems is the possibility to directly visualize local ordering using techniques such as magnetic force microscopy and Lorentz microscopy\cite{wang2006artificial}. A variety of interesting phenomena have been examined, such as emergent analogues for magnetic monopoles\cite{ladak2010direct}\cite{mengotti2011real}, vertex-based frustration\cite{gilbert2014emergent}\cite{morrison2013unhappy}, chiral dynamics\cite{gliga2017emergent}, and melting\cite{kapaklis2012melting}\cite{anghinolfi2015thermodynamic}\cite{sendetskyi2019continuous}. 

The ASI are typically composed of nanometre-sized ferromagnetic islands, designed to be single-domain and containing roughly $\sim10^5$ magnetic atoms. The island geometry provides a shape anisotropy that results in a preferred orientation for the single magnetic domains. In this way, Ising-like spin behaviour appears in that the magnetization of each island prefers to align along a single island determined axis. 
The first experimental realization of square ASI was done by Wang \emph{et al.}\cite{wang2006artificial} in 2006. Since then a number of other geometries have been studied including Kagome\cite{canals2016fragmentation, qi2008direct, farhan2014thermally, mengotti2011real, moller2009magnetic}, Shakti\cite{gilbert2014emergent}\cite{chern2013degeneracy}, and Triangle\cite{mol2012extending}\cite{rodrigues2013efficient}. Recently, experimental creation of 3D ASI systems\cite{perrin2016extensive}\cite{fernandez2017three}\cite{may2019realisation} has become possible which brings additional degrees of freedom, inspiring new theoretical studies\cite{moller2009magnetic}\cite{mol2010conditions}\cite{moller2006artificial}. It allows increased configurability and optimization of the magnetostatic interaction between different elements of the system.  One technique for fabricating 3D ASI uses two-photon laser lithography\cite{williams2018two}\cite{may2019realisation} through which it is possible to construct complex individual structures\cite{frenzel2017three}. Another approach with potential for large scale system is through layering different films. The idea in this case is to use lithography to pattern multilayers into 3D structures\cite{chern2014realizing}\cite{lavrijsen2013magnetic}.

\begin{figure}[h!]
\centering
\includegraphics[width=1.0\linewidth]{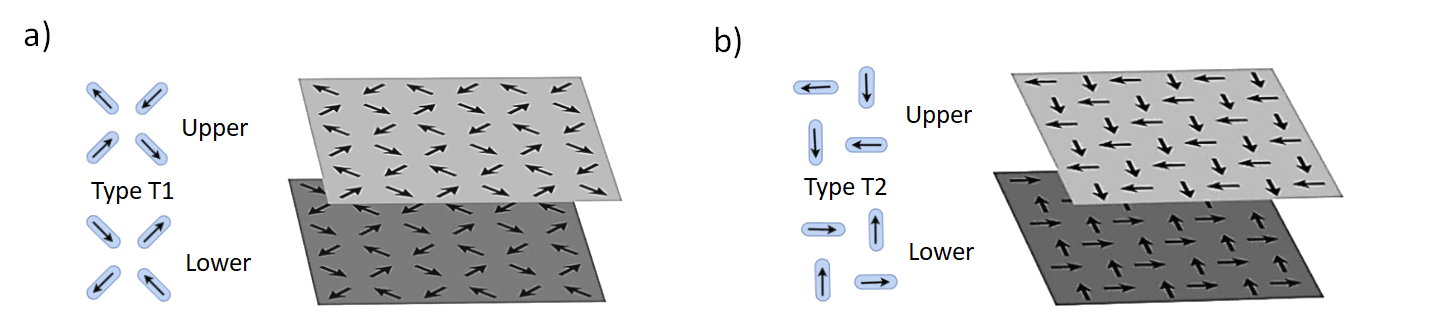}
\caption{Two types of bilayer ASI (BASI): a) a square and b) a pinwheel. In this example, each layer is composed of only T1 (for square) and T2 (for pinwheel) vertices. The lattice spacing in each layer is $a$, the distance between layers is $h$ and the angle $\phi$ defines the relative orientation of the two layers.}
\label{fig:fig1}
\end{figure}

 In this paper, we examine two different types of layered structures where the layers are rotated with respect to one another as sketched in Fig. \ref{fig:fig1}. The geometries of the individual layers are called square (Fig. \ref{fig:fig1}a)\cite{wang2006artificial} and pinwheel (Fig. \ref{fig:fig1}b)\cite{gliga2017emergent}\cite{macedo2018apparent}. Regardless of the geometry, the lattice for each layer has element spacing $a$ and the layers are separated by a distance $h$. The angle $\phi$ defines the relative orientation of the two layers. 

\begin{figure}[h!]
\centering
\includegraphics[width=0.8\linewidth]{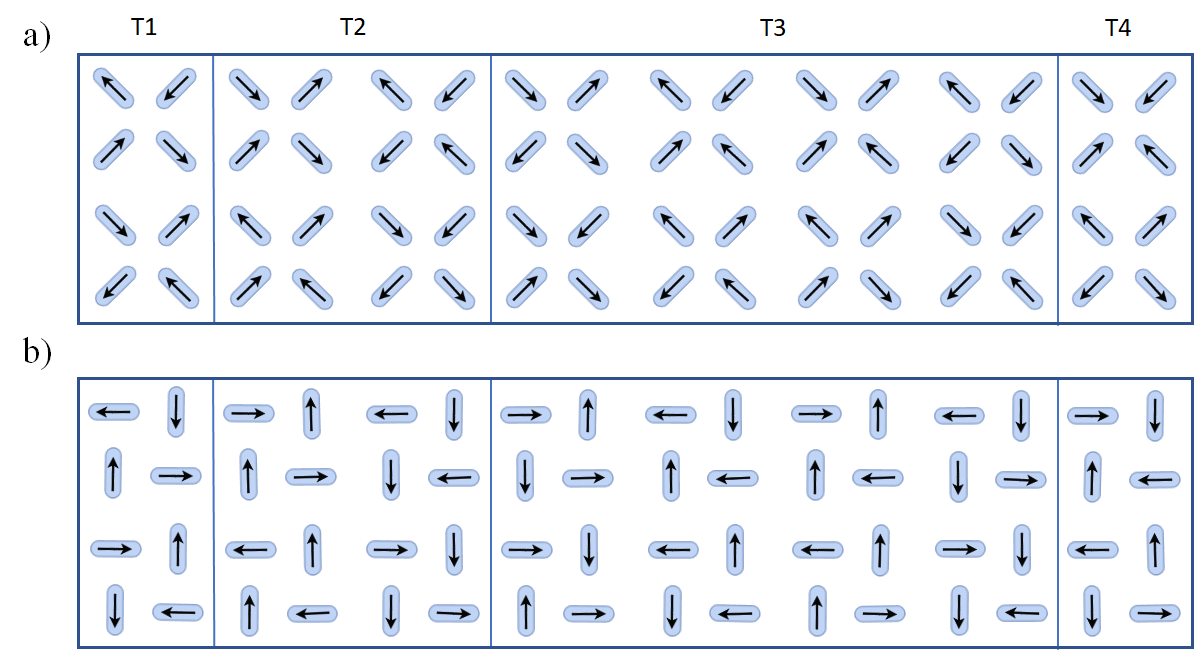}
\caption{(a) Sixteen possible vertex configurations in square artificial spin ice classified into four types. Vertex energy increases from left to right with the T1 being the lowest, followed by T2, then T3 and with T4 the highest. (b) Corresponding vertex types in pinwheel artificial spin ice. In this case, the lowest energy vertex is T2, followed by T3, then T1, with T4 again the highest.}
\label{fig:fig2}
\end{figure}

When the layers are sufficiently separated so as to not interact (i.e. $h \rightarrow\infty$), the square and pinwheel arrays have different ground state orderings. The orderings can be most easily described in terms of vertices. In the square and pinwheel ice, a vertex is defined as the meeting point of four neighbouring islands. These configurations are classified into four classes based on energy and are defined as T1, T2, T3, and T4 as sketched in Fig. \ref{fig:fig2}. Configurations within a given type have the same energy, i.e., they are degenerate. For the square geometry, type T1 has the lowest energy, is two-fold degenerate, and a tiling in one of the T1 configurations defines the ground state of a square lattice. These two type T1 subtypes are defined as T1(a) and T1(b). The three other classes have higher energies and are excitations above the ground state. The classification of the type of vertices in pinwheel geometry remains the same as for square ice, but now the lowest energy vertex is T2 and it is four-fold degenerate. These four degenerate subtypes are denoted as T2(a), T2(b), T2(c) and T2(d). Type T2(a) and T2(b) have opposite magnetization direction. The same goes for T2(c) and T2(d) pair. The other vertices are excitations above a ground state configured as a tiling of T2 vertices.

In this paper, we use numerical simulations to discuss how different values of $h$ and $\phi$ affect magnetic ordering under the influence of an applied external magnetic field for the BASI systems sketched in  Fig. \ref{fig:fig1}. We begin the discussion in the next section where we describe the model used for the simulations.

\section{Model And Methods} 
  
\begin{figure}[h!]
\centering
\includegraphics[width=0.9\linewidth]{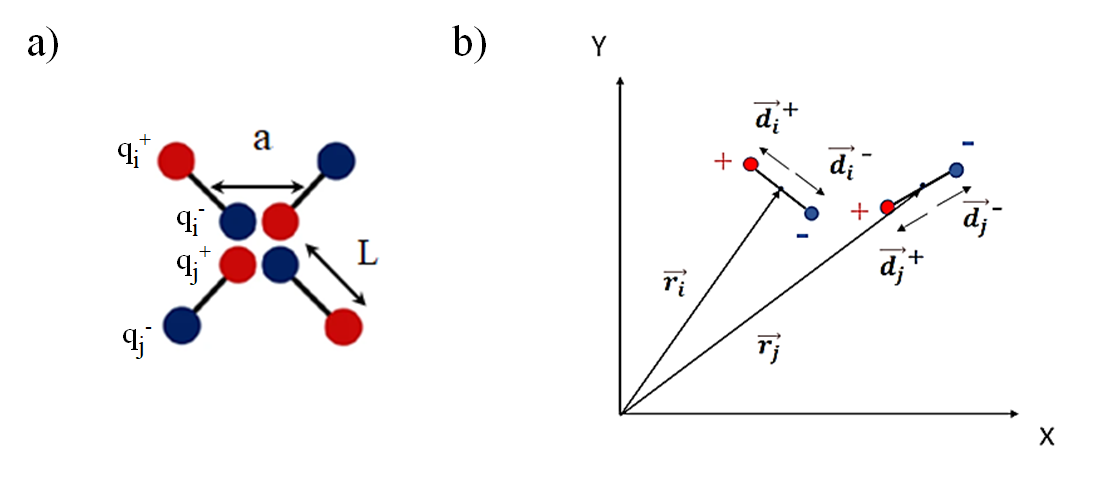}
\caption{Dumbbell model: Each island is approximated as a charged dumbbell of length $L$ in which two charges of equal magnitude but opposite signs sit at each end. In (a), a square vertex formed by four islands is shown with charges $\pm q_i$ at each site $i$. In (b) the vector location is shown for two dumbbells lying in the $xy$ plane. ${\vec{r}}_{i}$ and ${\vec{r}}_{j}$ are the vectors locating lattice sites $i$ and $j$. The $\vec{d}$'s have a magnitude that is half the length of the dumbbell i.e. $L/2$ and specifies the location of each magnetic charge relative to its associated lattice site.} 
\label{fig:fig4}
\end{figure}

The mutual interactions between the islands determine the collective behaviour of the system. To model this, we approximate each magnetic island as a charged dumbbell of length $L$ with magnetic charges of equal but opposite polarity sitting at either end as illustrated in Fig. \ref{fig:fig4}(a). The charges $\pm q_i$ are attached to lattice site $i$ and separated by a distance $L$ such that $\mu_s = qL$ where $\mu_s$ is the dipole moment of the spins, and $q$ is the magnitude of the magnetic charge. This dumbbell approximation is a powerful tool to understand the physics of artificial spin ice systems, specially to study the magnetic monopoles and their propagation through the ASI systems\cite{may2021magnetic}\cite{velo2020micromagnetic}. In this model, the ratio $L/a$ plays an important role: for example, M{\"o}ller and Moessner\citep{moller2006artificial} demonstrated that a larger value of the ratio $L/a$ results in the broadening of the temperature range for the ice rule obeying T1 and T2 vertices to form. In our work, we implement this charged dumbbell model since coupling between layers may be very sensitive to $L/a$, especially for small layer separations.

The positions of each charge are described by the vectors $\vec{r}$ and $\vec{d}$ as shown in Fig. \ref{fig:fig4}(b). The vectors $\vec{r}$ are taken with respect to the global cartesian reference frame with the origin located at the centre of the bottom layer. Subscripts on these vectors identify the lattice site within a layer such that the position of $+q_i$, for example, is given by $\vec{r_i}+\vec{d_i^{+}}$.

The configuration energy is computed as the sum of pairwise interactions of magnetic charges:

\begin{equation}
\begin{aligned}
\emph{E}_{ij} = {} & \emph{K} \Bigg[\frac{q_i^{+} q_j^{+}}{|({\vec{r}}_{i}+{\vec{d}}_{i}^{+}) - ({\vec{r}}_{j}+{\vec{d}}_{j}^{+})|} +\frac{q_i^{+} q_j^{-}}{|({\vec{r}}_{i}+{\vec{d}}_{i}^{+}) - ({\vec{r}}_{j}+{\vec{d}}_{j}^{-})|}\\
& +\frac{q_i^{-} q_i^{-}}{|({\vec{r}}_{i}+{\vec{d}}_{i}^{-}) - ({\vec{r}}_{j}+{\vec{d}}_{j}^{-})|}+\frac{q_i^{-} q_i^{+}}{|({\vec{r}}_{i}+{\vec{d}}_{i}^{-}) - ({\vec{r}}_{j}+{\vec{d}}_{j}^{+})|}\Bigg]
\end{aligned}
\label{eqn:eqn1}
\end{equation}

\noindent Each of the denominators represents a charge pair distance. Here, $\emph{K}$ is a dimensionless quantity defined as $\emph{K}=\dfrac{\mu_{0}}{{4\pi}{k_{B}}T}$ where $k_{B}$ is the Boltzmann constant and $T$ is the temperature. In what follows, the charges are assumed to interact strongly which is described by a dumbbell length $L = a/2$. The moment, $\mu_s$, is determined by the cross sectional area of the element end, and is assumed constant. The charge magnitude is then expressed as $q = \mu_s/L = \frac{\mu_s}{a/2}$.

Unless otherwise stated, each layer is a $20\times20$ array of elements for a total of $800$ moments in both layers. For a single ASI layer, the vectors in Eq. \ref{eqn:eqn1} run over directions $x$ and $y$, restricted to the layer plane. For the spins residing in the lower layer, $\vec{r_i}=(x_i,y_i,z_i=0)$ and for the upper layer spins, $\vec{r_i}=(x_i,y_i,z_i=h)$ where $h$ is the layer separation.

\section{Interlayer energy and alignment angle}

The energy of interaction between the layers depends on ($h$) and alignment angle ($\phi$). Consider the aligned case where $\phi=0$ for each geometry tiled with ground state vertices. As noted earlier, the ground states are two-fold degenerate for the square geometry which we denote here as `T1(a)' and `T1(b)'. When one layer is T1(a) and the other T1(b), the layered spins are anti-parallel for $\phi=0$ and the arrangement is called T1(a)-T1(b). Likewise, when the layers have the same ground state arrangement the layered spins are parallel. This configuration is called T1(a)-T1(a). For the pinwheel geometry, the anti-parallel, low energy type T2 subtypes T2(a)and T2(b) are chosen. The configurations T2(a)-T2(b) and T2(a)-T2(a) are explored.

We define the energy of interaction between layers, in the absence of an applied magnetic field,  as $V = E(\phi,h) - E_{\infty}$. Here, $E(\phi,h)$ is the energy of a BASI structure at layer separation $h$ and rotation angle, $\phi$, and $E_{\infty}$ is the energy for decoupled layers, i.e. the energy of the system when the layers are far apart ($h \rightarrow \infty$) so that they behave as isolated layers.  
 
\begin{figure}[h!]
\centering
\includegraphics[width=1.0\linewidth]{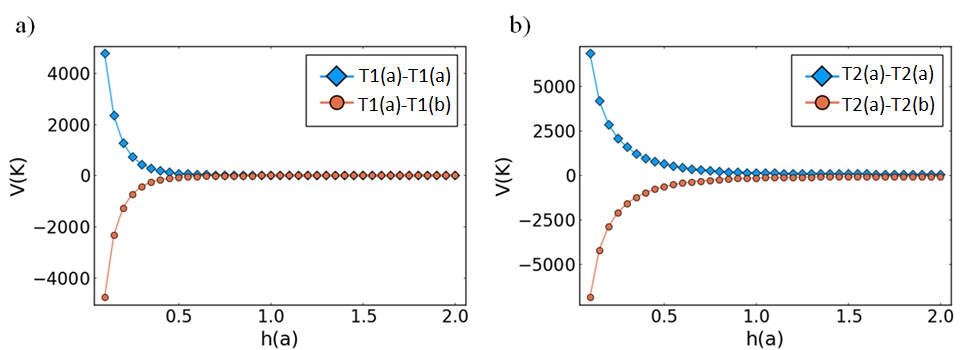}
\caption{Energy of interaction between the layers of a $\phi=0^{\circ}$ a) square and b) pinwheel BASI as a function of layer separation $h$. In the anti-parallel state, the layers experience a mutual attraction while for the parallel state, the potential is repulsive.}
\label{fig:fig7}
\end{figure}

First, we examine the interaction energy between the layers as a function of separation, $h$ with $\phi=0$.  The energy $V$ is given in units of $\emph{K}$ as a function of the separation $h$ (measured in units of lattice parameter a). Results are shown in Fig. \ref{fig:fig7}(a) for square ASI layers with T1 tiling and in (b) for the pinwheel geometry with T2 tiling. 

For the square geometry, when the layers have anti-parallel configuration (T1(a)-T1(b)), $V$ is strongly negative at very small height offsets ($h<<a$) and rapidly rises with increasing $h$ and vanishes as $h$ $>$ $a$. This is shown in Fig. \ref{fig:fig7}(a) where $V$ is plotted as a function of $h$. However, when in parallel (T1(a)-T1(a)) configuration, the opposite trend is observed with $V(\phi=0^{\circ},h)$ although still vanishing as $h$ $>$ $a$ (Fig. \ref{fig:fig7}(a)). When in anti-parallel (T1(a)-T1(b)) configuration, the layers attract each other while a mutual repulsion is experienced for T1(a)-T1(a) parallel state. These results for the squre ASI layers are in agreement with results from Nascimento \emph{et al.}\cite{nascimento2021bilayer} for no rotation ($\phi=0$).

Similar behaviour is observed for pinwheel BASI, except now the rate of change of the interaction potential between the layers is less than that for the square BASI geometry. This is shown in Fig. \ref{fig:fig7}(b).

\begin{figure}[h!]
\centering
\includegraphics[width=0.7\linewidth]{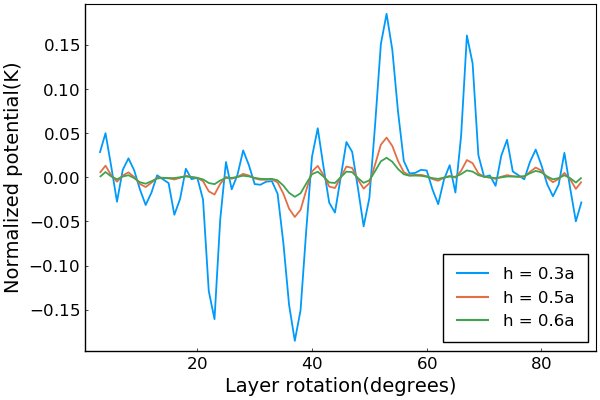}
\caption{Normalized potential energy of interaction between the layers of a square BASI structure (with T1(a)-T1(b) configuration) as a function of layer rotation at varying heights. There are distinctive peaks at certain angles. }
\label{fig:fig8}
\end{figure}

\begin{figure}[h!]
\centering
\includegraphics[width=0.7\linewidth]{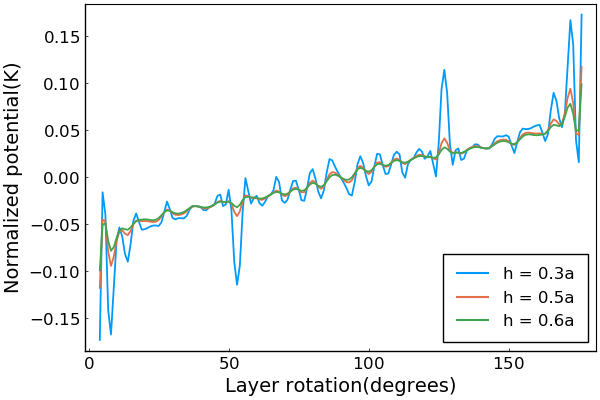}
\caption{Normalized potential energy of interaction between the layers of a pinwheel BASI structure (with T2(a)-T2(b) configuration) as a function of layer rotation at varying heights. }
\label{fig:fig9}
\end{figure}

The effects on $V$ for the misalignment of the layers ($\phi\ne 0$) are shown in Figs. \ref{fig:fig8} and \ref{fig:fig9}, for square and pinwheel geometries respectively. Here, $V$ is normalized to be unitless and the normalized potential per spin is plotted for $h=0.3a, 0.5a$ and $0.6a$ at different values of $\phi$. In both array geometries, distinct peaks are observed in $V$ at certain rotation angles. The closer the layers are, the sharper the peaks become. For the square array, the most significant peaks in $V(\phi,h)$  are observed at rotation angles $\phi$ = 38$^{\circ}$ and 54$^{\circ}$. Although not as large as in the square array, peaks are also noticeable in the pinwheel geometry at rotation angles $\phi$ = 54$^{\circ}$ and 128$^{\circ}$. These peaks originate from the competition between interlayer neighbour interactions. Since the square geometry has an antiferromagnetic ground state, the normalized potential baseline is flat. However, the ground state being ferromagnetic, the pinwheel BASI layers behave like two ferromagnets. Depending on the layer spacing, the contributions of the dipole moment, quadropole moment, and discrete spin elements dominate the characteristics of the potential plot.

\begin{figure}[h!]
\centering
\includegraphics[width=1.0\linewidth]{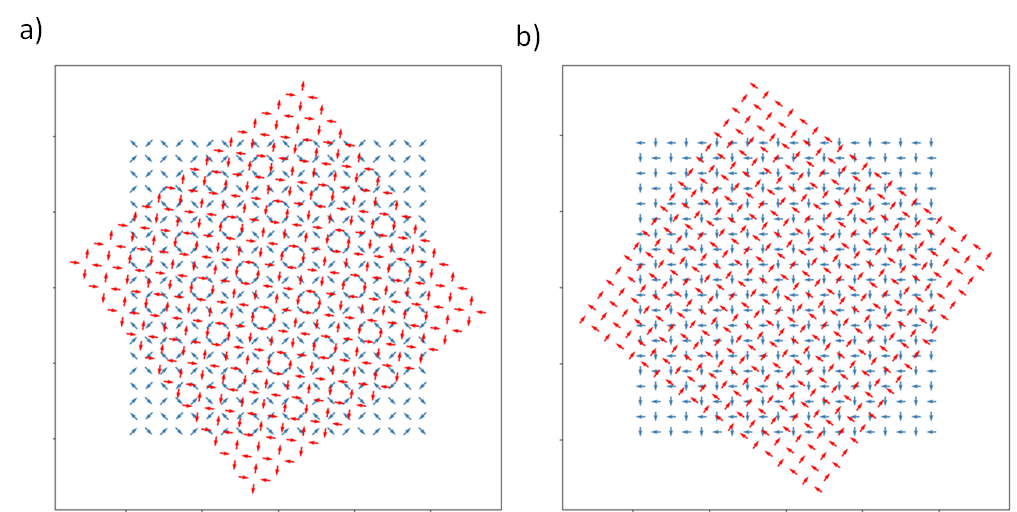}
\caption{Overlapping spin configurations in a) a square and b) a pinwheel BASI structure rotated at an angle $\phi$ = 38$^{\circ}$ and 54$^{\circ}$ respectively. Superlattice spin configurations appear for the bilayer structure.}
\label{fig:fig12}
\end{figure}

The peaks can be understood as follows. Certain rotation angles are special in that they define periodic arrangements of spins. Consider the geometries shown in Fig. \ref{fig:fig12} where anti-parallel configurations (T1(a)-T1(b) and T2(a)-T2(b)) are sketched at special angles in both geometries. The square geometry with rotation angle $\phi$ = 38$^{\circ}$ and the pinwheel geometry with $\phi$ = 54$^{\circ}$ exhibit a lateral `superlattice' structure where the two layers are superposed on one another. This structure consists of overlapping spins from adjacent layers with complex structure periodic on a scale larger than $a$, the lattice parameter. The structure is arranged in such a way that the ground state remains anti-ferromagnetic for square ice and ferromagnetic for pinwheel ice. The spin overlap contributes towards stronger interlayer coupling and, as discussed below, leads to interesting consequences for magnetization dynamics when driven by an applied magnetic field.

\section{Magnetic field driven dynamics} 

Magnetic ordering in the bilayer structures is affected by the presence of externally applied magnetic fields. Consider the magnetization reversal process in BASI under the influence of a spatially uniform external magnetic field applied parallel to the layers. In single layer square ASI, chain avalanche reversal is found\cite{morgan2011magnetic}\cite{morgan2013linear} whereas collective ferromagnetic behaviour leads to domain growth and shrinking in pinwheel single layer ASI \cite{li2018superferromagnetism}. 

\begin{figure}[h!]
\centering
\includegraphics[width=1.0\linewidth]{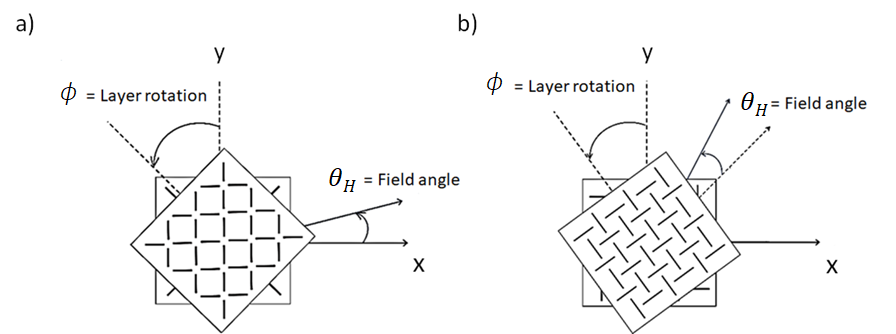}
\caption{Illustration of the layer rotation angle $\phi$ and field angle $\theta_H$ in a) square and b) pinwheel BASI. a) In square BASI, the applied field angle $\theta_H$ is measured with respect to the +$x$ axis. b) The diagonal to the +$x$ and +$y$ axes is used as the reference direction (black dotted arrow) for the field orientation in pinwheel BASI.}
\label{fig:fig13}
\end{figure}

With the application of a spatially uniform magnetic field $\mu_0 \vec{H}$, an additional energy term must be included when calculating the energy of the system: $\emph{E}_{ext} = - \mu_0 \sum_{i}\vec{\mu_{i}}\cdot \vec{H}$. Here, $\mu_{i}$ is the dipole moment of the $i$th dumbbell, and the sum is over all dumbbells in the bilayer system. The geometry for the applied field is defined in Fig. \ref{fig:fig13} for the square (a) and pinwheel (b) lattices. The applied field is in the $xy$ plane and makes an angle $\theta_H$ with respect to a symmetry axis of the bottom layer. For the square lattice, the reference axis is $x$. For the pinwheel lattice, the symmetry axis is taken as the $xy$ diagonal as this is one of the preferred ground state orientations of the ferromagnetic ordering.

In order to model the reversal dynamics instigated by the applied field, a self-consistent iterative algorithm is employed. The algorithm begins by choosing a direction and magnitude for the applied field, and then randomly choosing a lattice site on one of the layers.  At a chosen site $i$, the orientation of the moment $\vec{\mu}_i$ is chosen to minimize its energy. Note that in the dumbbell picture, a reorientation corresponds to changing the signs of the endpoint charges of that dumbbell. Keeping the field orientation and magnitude constant, the process is repeated for all lattice sites in the bilayer, each chosen in random order, until no reorientations occur. This usually occurs in less than $100$ iterations. The algorithm ensures that every lattice site is visited during each iteration. This configuration corresponds to a local minimum in the global energy for that field orientation and magnitude. The average number of vertices is then recorded. The field magnitude is then changed, and the iterative algorithm is repeated.  

Using this algorithm, a hysteresis loop is calculated by first setting a field large enough to align all spins as much as possible, then reducing the field in small steps until the spins align as much as possible along the reversed field direction. This process is repeated $20$ times for $20$ individual runs to get an average behaviour and the corresponding average magnetization as a function of the field value is studied. Also, in order to avoid trapping in metastable states due to high symmetry, the field is applied slightly away ($\sim 0.2^{\circ}$) from high symmetry directions. This contributes to the slight asymmetry in the hysteresis loops as we shall find out in the following sections. Both low and high field angles ($\theta_H$) are considered. We take the anticlockwise rotation direction to be positive.

\subsection{Square BASI}

Results of the simulation algorithm described above are presented here for superlattice angle ($\phi=38^{\circ}$) and unrotated ($\phi=0^{\circ}$) square BASI structures, induced by fields applied along directions $\theta_H = 0.2^{\circ}$ and $30.2^{\circ}$ offset from the +$x$ axis. The layers are assumed to be very close, separated by a distance $h=0.3a$ in order to produce strong interlayer coupling. The saturated high field spin configuration is polarized so that the tiling is T2 vertices in both layers aligned parallel to the magnetizing field for the initial saturated state.

\subsubsection{Square $\phi=0^{\circ}$ configuration}
\begin{figure}[H]
\centering
\includegraphics[width=0.6\linewidth]{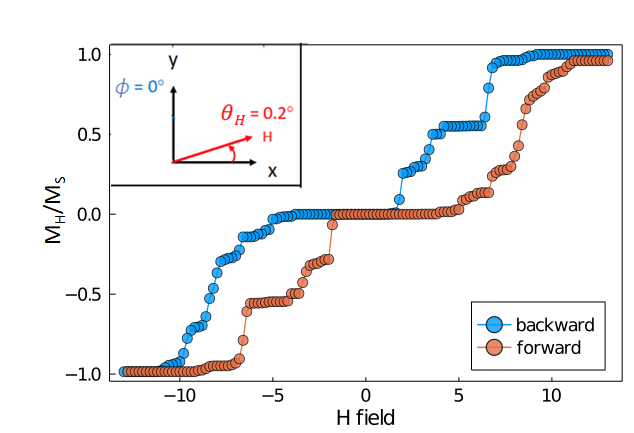}
\caption{Normalized net magnetization of a unrotated ($\phi=0$) square BASI configuration during a field sweep for $\theta_H=0.2^{\circ}$. The down (decreasing H) and up (increasing H) branches are indicated by the symbol color blue and red respectively. The inset illustrates the relative orientation of the applied field with respect to the +$x$ axis along with the layer rotation with respect to the +$y$ axis.}
\label{fig:fig15}
\end{figure}

In Fig. \ref{fig:fig15} we illustrate magnetization hysteresis loops obtained by taking averages of $20$ individual runs for square unrotated ($\phi=0$) configuration at small field angle $\theta_H=0.2^{\circ}$. The uncertainties are usually negligible ($<2\%$). The simulation is begun at a large field so that the array net moments are aligned with the +$x$ axis, and the magnetization is saturated. The field is decreased in strength in steps of $0.2$ in our reduced units until saturated magnetization is achieved along the -$x$ direction.

As seen in Fig. \ref{fig:fig15}, the overall structure is suggestive of a double loop with small nearly closed loops at large field magnitudes, and larger main loops at smaller fields. Starting at saturation, the magnetization remains constant until the field is reduced to $H=8.8$. At this point, the edge spins in both arrays start to flip as they are coupled to fewer neighbours than the bulk spins are.  

\begin{figure}[H]
\centering
\includegraphics[width=0.9\linewidth]{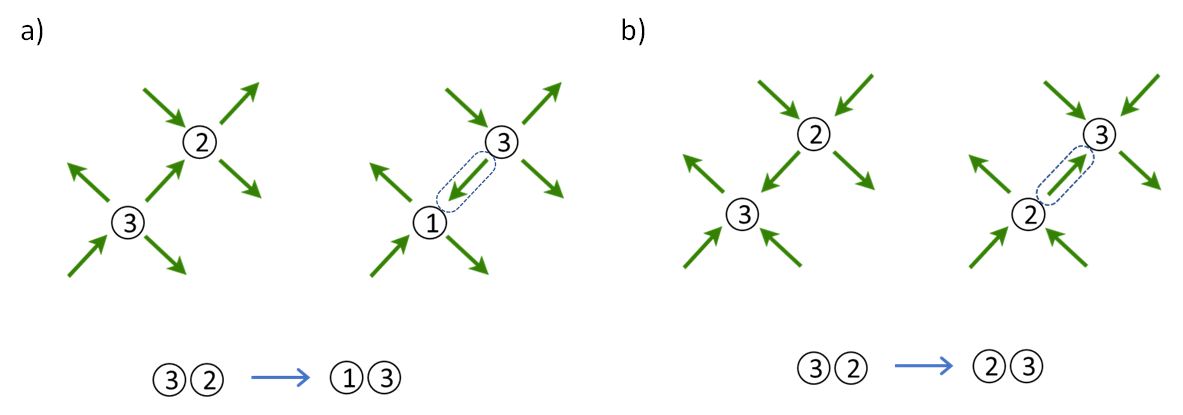}
\caption{T3 propagation process on a T2 background. a) T3 propagation with T1 creation. b) T3 propagation without T1 creation. The boxed spin is the one that has flipped. The notation $\textcircled{i}$ refers to a vertex of type $i$.}
\label{fig:fig17}
\end{figure}

Upon decreasing the field, a sharp drop in the magnetization is observed. This drop corresponds to the creation of T3 vertices along the array edges and their propagation in cascades of spin flips, leaving `trails' of T1 vertices through the array. Dynamics on a uniform T2 background always involve T3 vertex propagation because flipping a single spin in a T2 vertex always creates a T3 vertex. T3 propagation can occur in two processes: $\textcircled{3}\textcircled{2} \rightarrow \textcircled{1}\textcircled{3}$ and $\textcircled{3}\textcircled{2} \rightarrow \textcircled{2}\textcircled{3}$, which are shown in Fig. \ref{fig:fig17}. Here, we describe vertex processes using a notation where \textcircled{i} refers to a vertex of type $i$. So that a spin flip that converts a T3 - T2 pair into a T1 - T3 pair is written as $\textcircled{3}\textcircled{2} \rightarrow \textcircled{1}\textcircled{3}$.

\vspace{1em}

When the field angle $\theta_H$ is small (i.e. when the array net magnetization makes a relatively small angle with the field direction), the T3 propagation process $\textcircled{3}\textcircled{2} \rightarrow \textcircled{1}\textcircled{3}$ is favoured. This is because it is difficult to start dynamics if the array net magnetization is almost aligned with the field. Furthermore, since T1 vertices have the lowest energy, they are energetically favourable. As a result, T3 propagation via the creation of T1 vertices is energetically less expensive than $\textcircled{3}\textcircled{2} \rightarrow \textcircled{2}\textcircled{3}$ process.

\begin{figure}[H]
\centering
\includegraphics[width=0.8\linewidth]{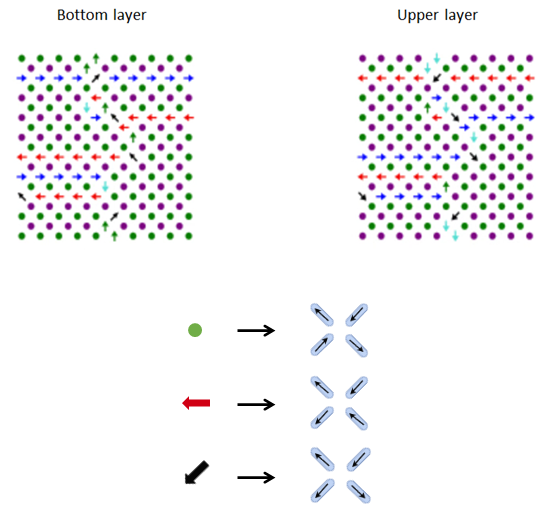}
\caption{Vertex configurations at applied field H = 0 for a $\phi=0^{\circ}$ square BASI configuration subject to a field applied at $0.2^{\circ}$ to the +$x$ axis. Colour-coded arrows represent type T2 and T3 vertex magnetization orientations. The circles represent alternating types of type T1 vertices. Type T4 vertices are not observed. At the bottom, the spin arrangements corresponding to the vertex magnetization orientations are presented for one from each type. Note that for this field with aligned layers, the layer spins are perfectly antiparallel with respect to one another throughout the avalanche vertex processes.}
\label{fig:fig18}
\end{figure}

As the field is decreased further, other types of vertex processes appear, and chains/avalanches are blocked. For example, the appearance of T3 vertices in the bulk via the process $\textcircled{2}\textcircled{2} \rightarrow \textcircled{3}\textcircled{3}$. Also, the field being weak, interlayer coupling slowly takes over and favours certain configurations where the spins in the adjacent layers try to achieve a local anti-parallel state. These are metastable states and give rise to a plateau-like feature where the net magnetization remains unchanged for a range of field values. The largest plateau corresponds to magnetization $M_H/M_S=0$ as seen in Fig. \ref{fig:fig15}. As the field becomes sufficiently small, the interlayer coupling forces the vertex moments in the adjacent layers to achieve a complete anti-parallel configuration, and the net magnetization vanishes. Hence, we do not observe any remnant magnetization at $0$ field, although individual arrays contain a small net moment. The corresponding vertex configurations for one of the sample runs are shown in Fig. \ref{fig:fig18}. The full reversal process along with snapshots of vertex configurations at some certain field values can be found in the appendix.

As the field is decreased further, a slow evolution towards the completely reversed state is observed. Most significantly, the corresponding spins on the two separate layers remain anti-parallel and track each other perfectly during avalanche reversal processes.

\begin{figure}[h!]
\centering
\includegraphics[width=0.6\linewidth]{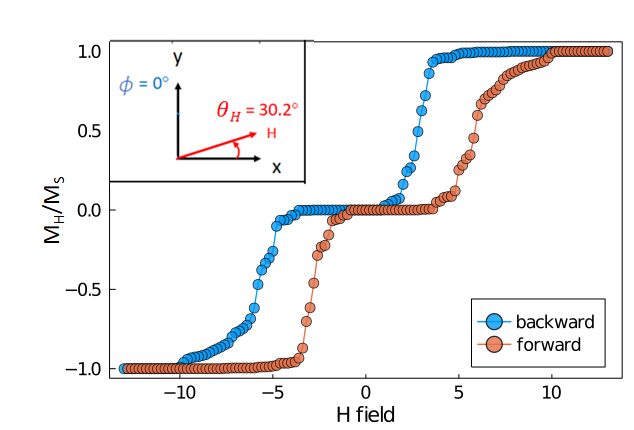}
\caption{Normalized net magnetization of a unrotated ($\phi=0^{\circ}$) square BASI configuration during a field sweep for $\theta_H=30.2^{\circ}$. The down (decreasing H) and up (increasing H) branches are indicated by the symbol color blue and red respectively. Inset shows the relative orientation of the applied field with respect to the +$x$ axis along with the layer rotation with respect to the +$y$ axis.}
\label{fig:fig20}
\end{figure}

The double loop structure seen in Fig. \ref{fig:fig15}  is found only for relatively small $\theta_H$. At a larger angle $\theta_H=30.2^{\circ}$ with respect to the $+x$ axis, only the two small field main loops appear. An example hysteresis is shown in Fig. \ref{fig:fig20}.

\begin{figure}[h!]
\centering
\includegraphics[width=0.9\linewidth]{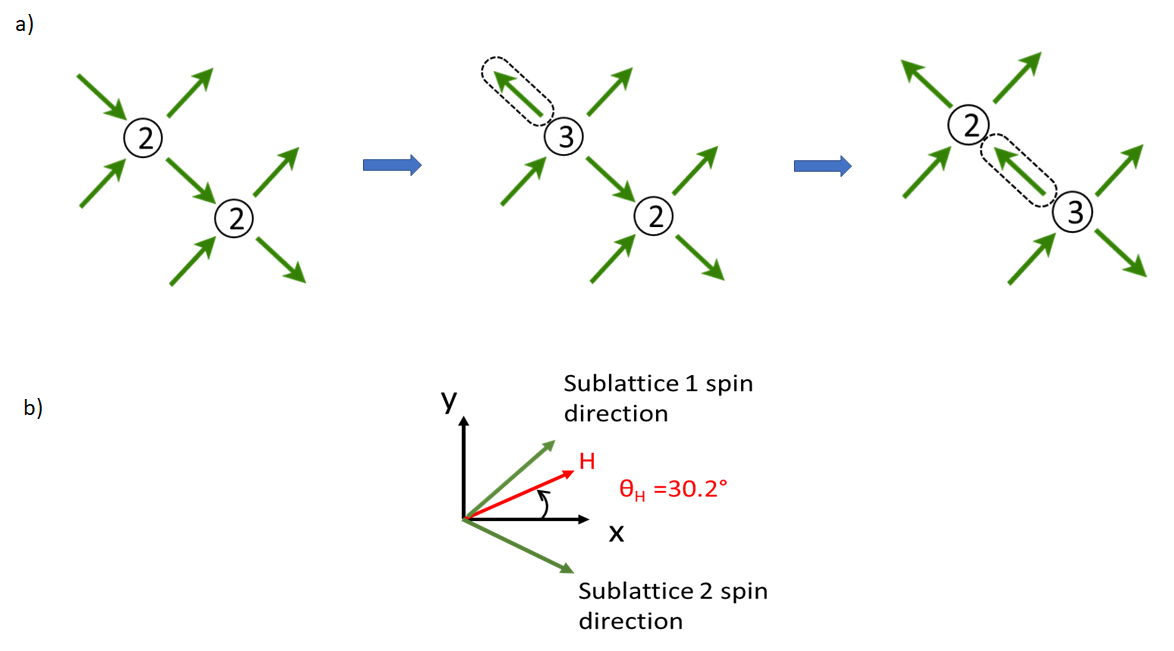}
\caption{a) T3 propagation via vertex process $\textcircled{3}\textcircled{2} \rightarrow \textcircled{2}\textcircled{3}$. The notation $\textcircled{i}$ refers to a vertex of type $i$. The boxed spin is the one that has flipped. b) The relative angles between the applied field and the array sublattices. It can be clearly seen that sublattice 2 makes a larger angle with the field. Hence the spin on that sublattice is the one that has flipped and converted a $\textcircled{3}\textcircled{2}$ vertex pair to a $\textcircled{2}\textcircled{3}$ vertex pair.}
\label{fig:fig21}
\end{figure}

\begin{figure}[H]
\centering
\includegraphics[width=0.8\linewidth]{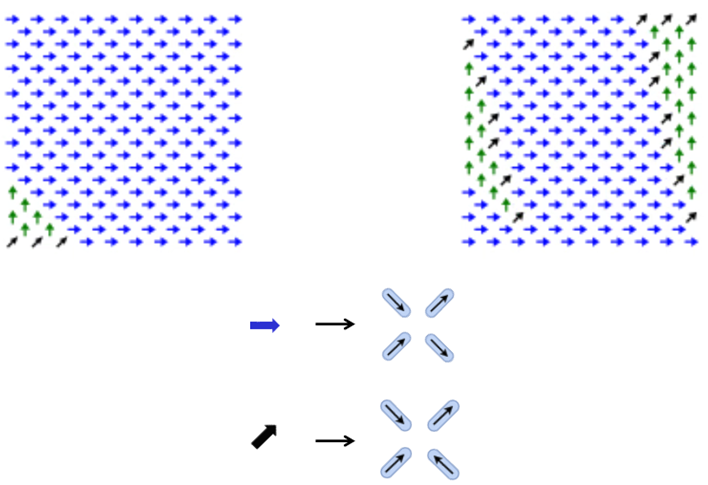}
\caption{Vertex configurations at applied field H = 6.8 for a $\phi=0^{\circ}$ square BASI configuration subject to a field applied at $\theta_H=30.2^{\circ}$ to the +$x$ axis. Unlike the small $\theta_H$ case, for an angle of $\theta_H=30.2^{\circ}$ the spins in the two layers do not track one another. Colour-coded arrows represent type T2 and T3 vertex magnetization orientations. Type T3 vertex propagation via process $\textcircled{3}\textcircled{2} \rightarrow \textcircled{2}\textcircled{3}$ can be observed in the upper layer. At the bottom, the spin arrangements corresponding to the vertex magnetization orientations are presented for one from each type.}
\label{fig:fig22}
\end{figure}

Spins on one of the array sublattices make a much larger angle with the applied field, and as sketched in Fig. \ref{fig:fig21}, this results in T3 vertex propagation via $\textcircled{3}\textcircled{2} \rightarrow \textcircled{2}\textcircled{3}$ process. Example vertex configurations for one of the sample runs at $H=6.8$ are shown in Fig. \ref{fig:fig22} for each layer. In this case processes and spins in the separate layers do not track one another despite strong interlayer coupling due to the close proximity of the layers. The full reversal process along with snapshots of vertex configurations at some certain field values can be found in the appendix.

\subsubsection{Square $\phi=38^{\circ}$ configuration}

\begin{figure}[h!]
\centering
\includegraphics[width=0.6\linewidth]{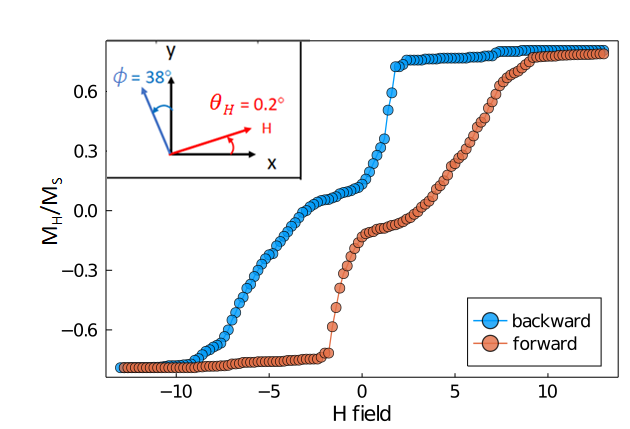}
\caption{Normalized net magnetization of a rotated ($\phi=38^{\circ}$) square BASI configuration during a field sweep for $\theta_H=0.2^{\circ}$. The down (decreasing H) and up (increasing H) branches are indicated by the symbol color blue and red respectively. Inset shows the relative orientation of the applied field with respect to the +$x$ axis along with the layer rotation with respect to the +$y$ axis.}
\label{fig:fig23}
\end{figure}

\begin{figure}[H]
\centering
\includegraphics[width=0.8\linewidth]{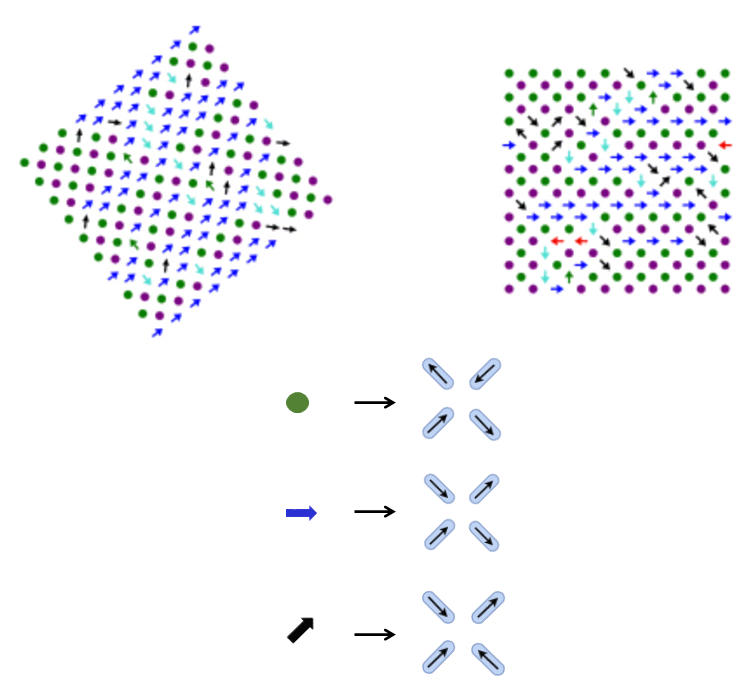}
\caption{Vertex configurations at applied field H = 0.6 for a $\phi=38^{\circ}$ square BASI configuration subject to a field applied at $0.2^{\circ}$ to the +$x$ axis. Colour-coded arrows represent type T2 and T3 vertex magnetization orientations. The circles represent alternating types of type T1 vertices. Type T4 vertices are not observed. At the bottom, the spin arrangements corresponding to the vertex magnetization orientations are presented for one from each type.}
\label{fig:fig25}
\end{figure}

Figure \ref{fig:fig23} shows magnetization hysteresis loops obtained by taking averages of $20$ individual runs for a layer rotation of $\phi=38^{\circ}$  at a small field angle $\theta_H=0.2^{\circ}$. The uncertainties are negligible ($<2\%$). As before, hysteresis begins as a large field with saturated magnetization. As seen in Fig. \ref{fig:fig23}, the magnetization remains constant until the field is reduced to $H=4.2$. The net magnetization in the bottom array initially makes an angle of $37.8^{\circ}$ with the applied field whereas, for the upper array, the magnetization is aligned as near parallel to the field as possible. Unlike the unrotated ($\phi=0^{\circ}$) configuration, several vertex processes are now involved including the two T3 vertex propagation processes ($\textcircled{3}\textcircled{2} \rightarrow \textcircled{1}\textcircled{3}$ and $\textcircled{3}\textcircled{2} \rightarrow \textcircled{2}\textcircled{3}$). This leads to complex configurations such as those shown in Fig. \ref{fig:fig25} for one of those individual runs. Note that processes in each layer are very different from one another.

Because of the relative angle between the layers, interlayer coupling fails to force the vertex moments in the adjacent layers to achieve a complete antiparallel alignment. There are consequently no plateaus in the magnetization curves shown in Fig. \ref{fig:fig23}. At $H = 0$, the array moments do not cancel each other completely, and there is a remnant magnetization at  $\theta_H=0.2^{\circ}$. As the field increases in the negative direction, the magnetization increases smoothly until it jumps to saturation. The full reversal process along with snapshots of vertex configurations at some certain field values for one of the sample runs are included in the appendix.

At a large field angle when $\theta_H=30.2^{\circ}$, the $H$ field at which spin reversals occur and the dynamics begin is the same as for other field orientations. The magnetization reversal process is also similar.

\subsection{Pinwheel BASI}
\label{sub:subsection6} 
In this section, magnetization reversal dynamics are examined for superlattice angle ($\phi=54^{\circ}$) and unrotated ($\phi=0^{\circ}$) pinwheel BASI structures, induced by fields applied along directions $\theta_H =$ 0.2$^{\circ}$ and 30.2$^{\circ}$ offset from the diagonal to the $+x$ and $+y$ axes. Note that single layer pinwheel ASI has a ferromagnetic ground state\cite{macedo2018apparent}, and effects from stray fields produced at the array edges will strongly affect the resulting magnetization processes. The following results were obtained with separation $h=0.5a$ which reduces this effect for the size of the arrays considered.

\subsubsection{Pinwheel $\phi=0^{\circ}$ configuration}

\begin{figure}[h!]
\centering
\includegraphics[width=0.6\linewidth]{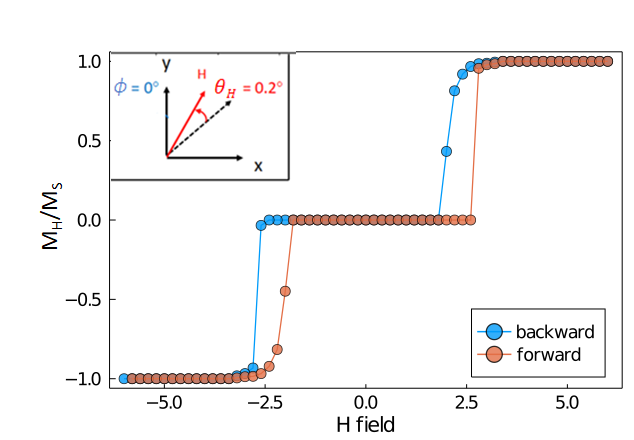}
\caption{Normalized net magnetization of a unrotated ($\phi=0^{\circ}$) pinwheel BASI configuration during a field sweep for $\theta_H=0.2^{\circ}$. The down (decreasing H) and up (increasing H) branches are indicated by the symbol color blue and red respectively. The inset shows the relative orientation of the applied field with respect to the $xy$ diagonal along with the layer rotation with respect to the +$y$ axis.}
\label{fig:fig29}
\end{figure}

Magnetization hysteresis results obtained by taking averages of $20$ individual runs are shown in Fig. \ref{fig:fig29} for unrotated pinwheel layers ($\phi=0^{\circ}$)  at a small field angle $\theta_H=0.2^{\circ}$. The uncertainties are usually negligible ($<2\%$). The simulation is begun as before with a  large field that saturates the magnetization of both layers along the diagonal to the +$x$ and +$y$ axes. The field is then decreased in strength in steps of 0.2 in reduced units and then increased in the reverse direction until saturated magnetization is again achieved.

\begin{figure}[h!]
\centering
\includegraphics[width=0.7\linewidth]{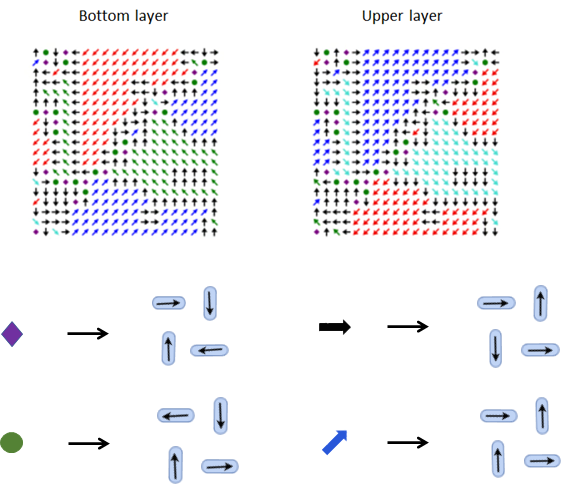}
\caption{Vertex configurations at applied field H = 0 for a $\phi=0^{\circ}$ pinwheel BASI configuration subject to a field applied at an angle $0.2^{\circ}$ to the $xy$ diagonal. Colour-coded arrows represent type T2 and T3 vertex magnetization orientations. The circles represent T1 vertices while T4 vertices are indicated by the diamond symbols. At the bottom, the spin arrangements corresponding to the vertex magnetization orientations are presented for one from each type.}
\label{fig:fig31}
\end{figure}

As seen in Fig. \ref{fig:fig29}, the magnetization remains constant until the field is reduced to $H=3.4$. Domains of different shapes and sizes appear in the spin configurations of each layer in response to stray fields from the array edges. At sufficiently small fields, spins in the adjacent layers align antiparallel, and the net magnetization vanishes. This transition is signaled by the sharp drops in $M_H/M_S$ in Fig. \ref{fig:fig29} at a field around $H=2.5$. The system remains in this stable configuration for a range of field values.  The vertex configurations at $H=0$ for one of the sample runs are shown in Fig. \ref{fig:fig31}. The full reversal process along with snapshots of vertex configurations at some certain field values are included in the appendix.

\begin{figure}[h!]
\centering
\includegraphics[width=0.6\linewidth]{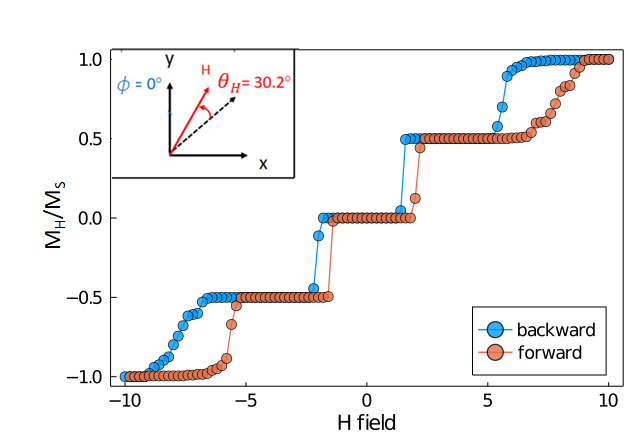}
\caption{Normalized net magnetization of a unrotated ($\phi=0^{\circ}$) pinwheel BASI configuration during a field sweep for $\theta_H=30.2^{\circ}$. The down (decreasing H) and up (increasing H) branches are indicated by the symbol color blue and red respectively. Inset shows the relative orientation of the applied field with respect to the $xy$ diagonal along with the layer rotation with respect to the +$y$ axis.}
\label{fig:fig33}
\end{figure}

\begin{figure}[h!]
\centering
\includegraphics[width=1.0\linewidth]{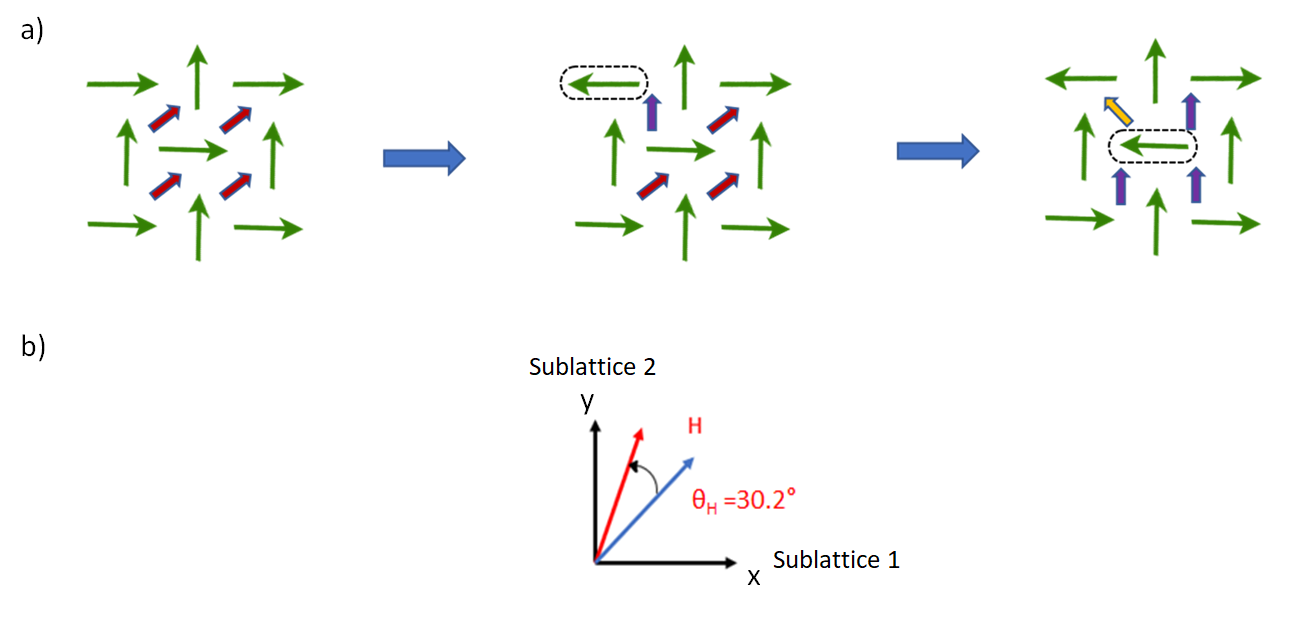}
\caption{a) Vertex process $\textcircled{3}\textcircled{2} \rightarrow \textcircled{2}\textcircled{3}$ in a unrotated ($\phi=0^{\circ}$) pinwheel BASI configuration during a field sweep for $\theta_H=30.2^{\circ}$. The notation $\textcircled{i}$ refers to a vertex of type $i$. The boxed spin is the one that has flipped. b) The relative angles between the applied field and the array sublattices. It can be clearly seen that sublattice 1 makes a larger angle with the field. Hence the spins on that sublattice are the ones that have flipped.}
\label{fig:fig34}
\end{figure}

A very different double loop hysteretic behaviour is observed when the field is applied at a larger angle. An example for $\theta_H=30.2^{\circ}$ relative to the diagonal in the $xy$ plane is shown in  Fig. \ref{fig:fig33}. Spins on one of the array sublattices make a much larger angle with the applied field than on the other sublattice. For this angle, the vertex process $\textcircled{3}\textcircled{2} \rightarrow \textcircled{2}\textcircled{3}$ is favoured in both arrays. A sketch of how spins reverse in this process is shown in Fig. \ref{fig:fig34}.

\begin{figure}[h!]
\centering
\includegraphics[width=0.7\linewidth]{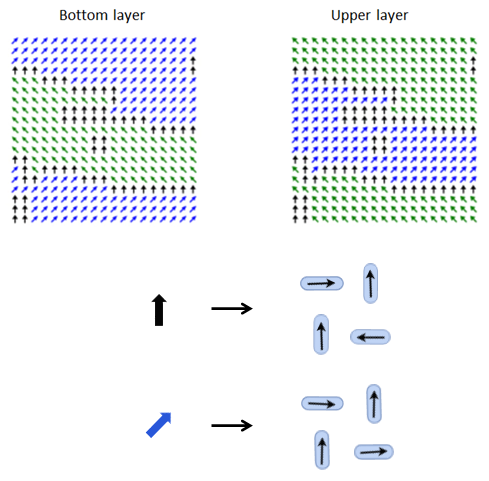}
\caption{Vertex configurations at applied field $H=4$ for a $\phi=0^{\circ}$ pinwheel BASI configuration subject to a field applied at an angle $30.2^{\circ}$ to the $xy$ diagonal. Colour-coded arrows represent type T2 and T3 vertex magnetization orientations. At the bottom, the spin arrangements corresponding to the vertex magnetization orientations are presented for one from each type.}
\label{fig:fig35}
\end{figure}

The initial reversal starts at one of the corners (perpendicular to the field direction) and leads to T2 domain formation with a domain wall consisting of T3 vertices. The domain expands in size gradually, and reversals begin to appear at other locations. A stable configuration is reached where spins on one of the sublattices align anti-parallel to their nearest neighbours on the other layer. This results in the magnetization plateau at $M_H/M_S=0.5$ (see Fig. \ref{fig:fig33}). The associated vertex configurations are shown in Fig. \ref{fig:fig35}.

With the decreasing field, small domains form and interlayer coupling forces the vertex moments in the adjacent layers to achieve a complete anti-parallel configuration, and $M_H/M_S$ becomes zero and remains for small magnitude fields. As the field becomes larger in the negative direction, the spins evolve back towards saturation with another plateau at larger fields. The full reversal process along with snapshots of vertex configurations for one of the sample runs can be found in the appendix.

\subsubsection{Pinwheel $\phi=54^{\circ}$ configuration}

\label{sub:subsubsection4}

\begin{figure}[H]
\centering
\includegraphics[width=0.6\linewidth]{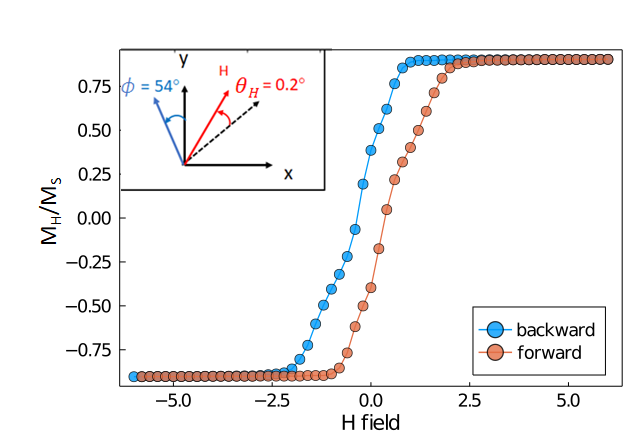}
\caption{Normalized net magnetization of a rotated ($\phi=54^{\circ}$) pinwheel BASI configuration during a field sweep for $\theta_H=0.2^{\circ}$. The down (decreasing H) and up (increasing H) branches are indicated by the symbol color blue and red respectively. Inset shows the relative orientation of the applied field with respect to the $xy$ diagonal along with the layer rotation with respect to the +$y$ axis.}
\label{fig:fig36}
\end{figure}

The hysteresis shown in Fig. \ref{fig:fig36} is for a pinwheel rotated lattice at the superlattice angle $\phi=54^{\circ}$ with respect to the other pinwheel lattice with a field applied at $\theta_H=0.2^{\circ}$. The results are obtained by taking averages of $20$ individual runs. The uncertainties are negligible ($<2\%$). Similar hysteresis loops are found for the field applied at larger angles.

\begin{figure}[H]
\centering
\includegraphics[width=0.6\linewidth]{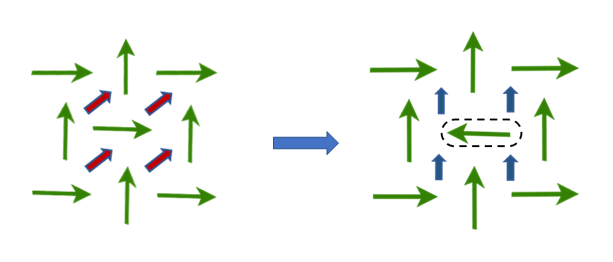}
\caption{Type T3 vertex creation in bulk in double pairs. Only a single spin flip is needed for this. The boxed spin is the one that has flipped.}
\label{fig:fig39}
\end{figure}

The spin reversal dynamics begins at $H\approx 1.2$ for all field orientations, and it takes place in the regions where the layers do not overlap. Moreover, the initial reversals are suppressed in the layer with the field aligned along the T2 moment direction and begin instead in the other layer where alignment of the T2 vertex moments is less favourable. Initially, T3 vertices are created around the edges and propagated via the process 
$\textcircled{3}\textcircled{2} \rightarrow \textcircled{2}\textcircled{3}$.  However, as the field approaches $H\approx 1.2$, T3 vertices are created in bulk as well in double pairs (see Fig. \ref{fig:fig39}).

\begin{figure}[H]
\centering
\includegraphics[width=0.7\linewidth]{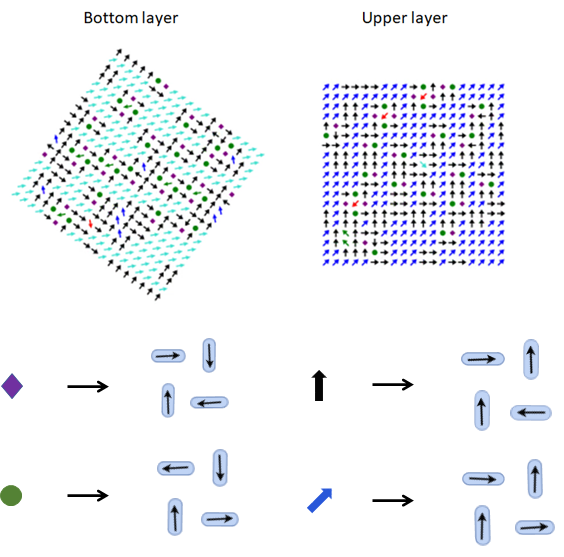}
\caption{Vertex configurations at applied field $H=0.4$ for a $\phi=54^{\circ}$ pinwheel BASI configuration subject to a field applied at an angle $0.2^{\circ}$ to the $xy$ diagonal. Colour-coded arrows represent type T2 and T3 vertex magnetization orientations. The circles represent T1 vertices while T4 vertices are indicated by the diamond symbols. At the bottom, the spin arrangements corresponding to the vertex magnetization orientations are presented for one from each type.}
\label{fig:fig40}
\end{figure}

As the field is reduced in strength, other vertex processes start to occur and a complex configuration is achieved. The corresponding vertex configurations are shown in Fig. \ref{fig:fig40} for one of the sample runs. It should be mentioned that the reversal process that takes place in this superlattice angle geometry is significantly different from the unrotated one. Here, clear domain formation and domain wall propagation are not observed, unlike the unrotated pinwheel geometry (see Fig. \ref{fig:fig40}). The full reversal process can be found in the appendix.

Due to the geometrical constraints, anti-parallel alignment of the vertex moments in adjacent layers is not possible. As a consequence, a remnant magnetization at $H = 0$ exists. Reversal occurs smoothly without the presence of any plateau-like feature.

\section{Conclusion}
The magnetization reversal processes in bilayer artificial spin ice systems (BASI) have been studied using numerical simulations for square and pinwheel geometries. The relative orientation of each layer is varied and shown to strongly affect reversal processes. Moreover, at certain angles, a superlattice structure appears which, when the layers are strongly coupled, leads to complex reversal dynamics that are mirrored in each layer. Interestingly, both square and pinwheel geometries exhibit a very different hysteretic behaviour at the superlattice angle when compared to other angles. Also, changing the orientation of the applied field has less effect on reversal for the superlattice angles regardless of the geometry. 

As in isolated layers, reversal in the square geometry occurs via T2 chain avalanche, whereas domain growth and shrinking drive reversal processes in pinwheel geometry. Square rotated and unrotated configurations reverse differently under the influence of an external field. Zero field remnant magnetization is absent in unrotated square BASI but is present in configuration with superlattice angle. Also, the magnetization curve for unrotated ($\phi=0^{\circ}$) square BASI exhibits plateau-like features and sharp drops in magnetization. On the other hand, for rotated ($\phi=38^{\circ}$) square BASI, it is relatively smoother with no plateau-like features. 

A similar trend is observed in pinwheel geometry. Magnetization curves for pinwheel rotated and unrotated configurations display different features. Unlike unrotated ($\phi=0^{\circ}$) pinwheel BASI, the superlattice angle ($\phi=54^{\circ}$) exhibits a remnant magnetization at zero field. Plateaus can be seen in the magnetization curve for $\phi=0^{\circ}$ pinwheel BASI. However, this is not the case for the rotated ($\phi=54^{\circ}$) structure. Also, reversal processes in the unrotated and superlattice angle ($\phi=54^{\circ}$) pinwheel geometry are quite distinctive. While unrotated one exhibits the formation of domains of different shapes and sizes along with domain wall motion during the process, the $\phi=54^{\circ}$ rotated geometry shows no such behaviour. Instead, patterns with different shapes are observed.

\begin{acknowledgements}
This work was supported by The Natural Sciences and Engineering Research Council of Canada (NSERC) Discovery, John R. Leaders Fund - Canada Foundation for Innovation (CFI-JELF), Research Manitoba and the University of Manitoba, Canada.
\end{acknowledgements}

\bibliographystyle{unsrt}
\bibliography{references-fixed}

\section{APPENDIX}
\subsection{Magnetization reversal process in square BASI}
\label{APP:config} 
\begin{figure}[h!]
\centering
\includegraphics[width=1.0\linewidth]{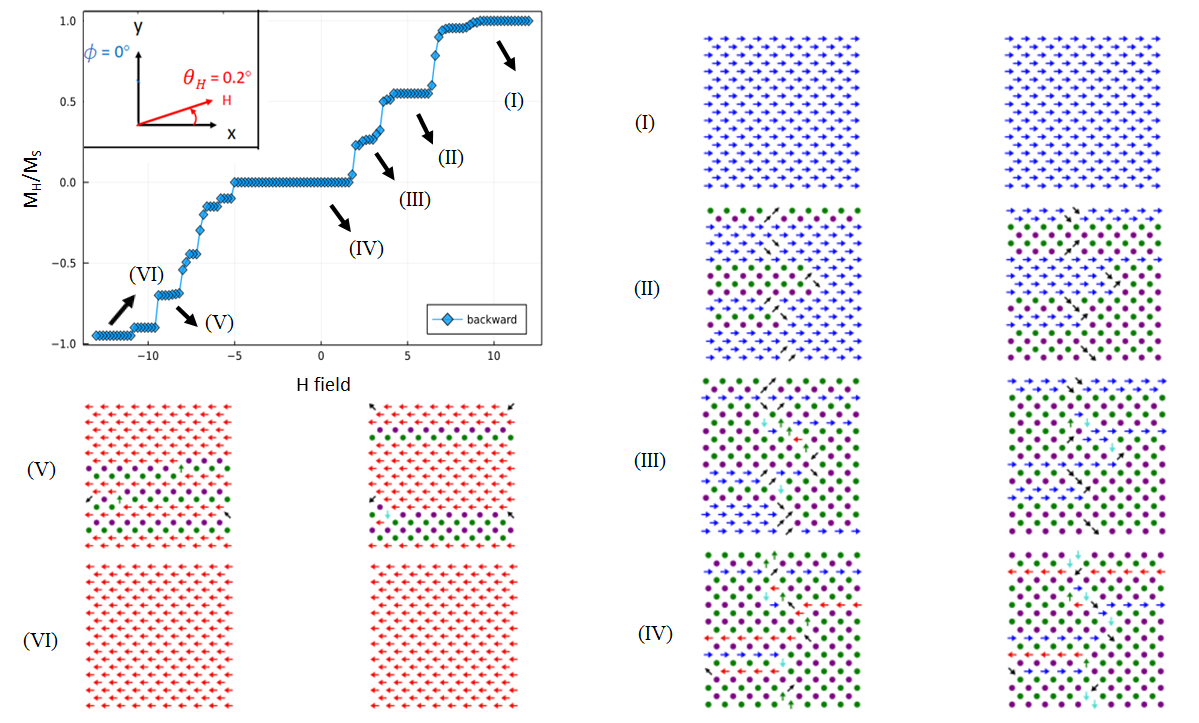}
\caption{Magnetization reversal process in a $\phi=0^{\circ}$ square BASI configuration subject to a field applied at an angle $0.2^{\circ}$ to the +$x$ axis. Six points are marked in the hysteresis loop across the reversal. The snapshots of the vertex configurations at each of these marked points are also given. Colour-coded arrows represent type T2 and T3 vertex magnetization orientations. The circles represent alternating types of type T1 vertices. Type T4 vertices are not observed.}
\label{fig:fig41}
\end{figure}

\begin{figure}[h!]
\centering
\includegraphics[width=1.0\linewidth]{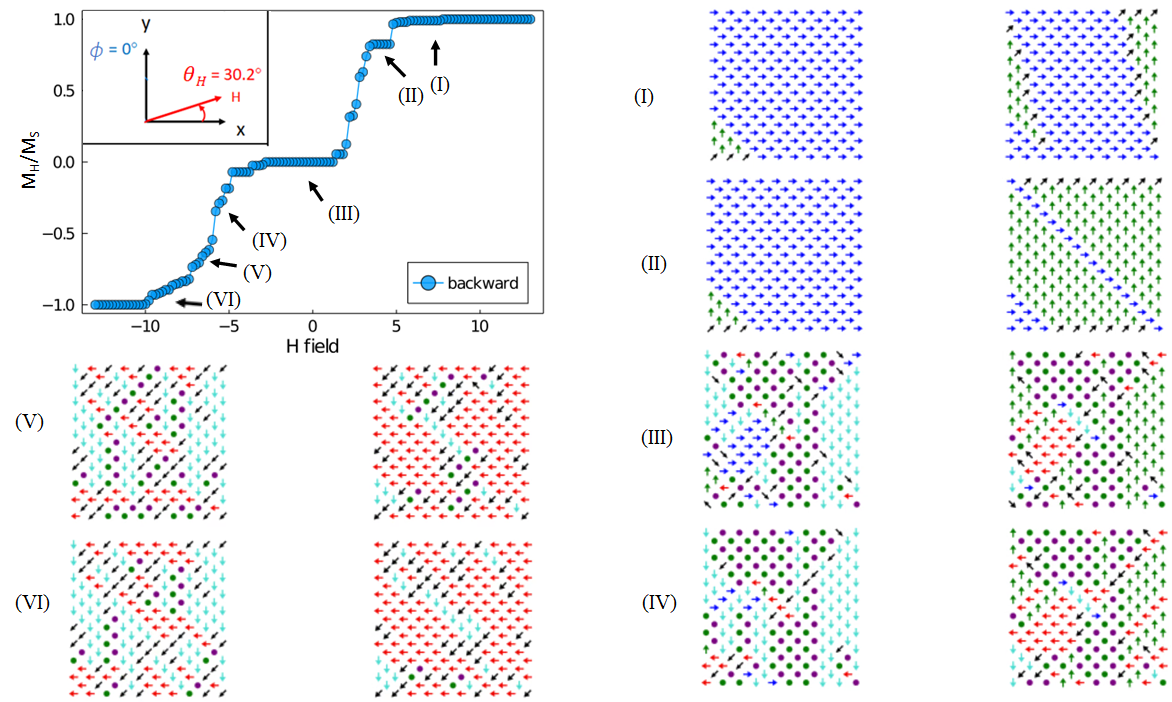}
\caption{Magnetization reversal process in a $\phi=0^{\circ}$ square BASI configuration subject to a field applied at an angle $30.2^{\circ}$ to the +$x$ axis. Six points are marked in the hysteresis loop across the reversal. The snapshots of the vertex configurations at each of these marked points are also given. Colour-coded arrows represent type T2 and T3 vertex magnetization orientations. The circles represent alternating types of type T1 vertices. Type T4 vertices are not observed.}
\label{fig:fig42}
\end{figure}

\begin{figure}[H]
\centering
\includegraphics[width=1.0\linewidth]{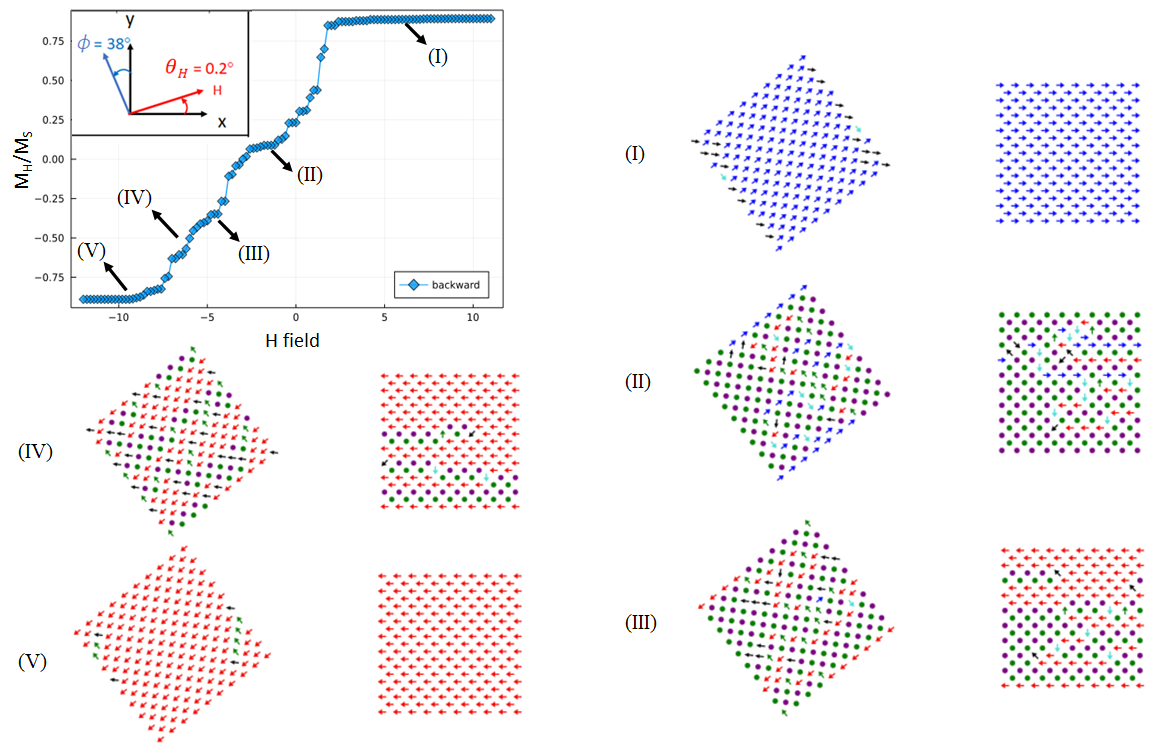}
\caption{Magnetization reversal process in a $\phi=38^{\circ}$ square BASI configuration subject to a field applied at an angle $0.2^{\circ}$ to the +$x$ axis. Five points are marked in the hysteresis loop across the reversal. The snapshots of the vertex configurations at each of these marked points are also given. Colour-coded arrows represent type T2 and T3 vertex magnetization orientations. The circles represent alternating types of type T1 vertices. Type T4 vertices are not observed.}
\label{fig:fig43}
\end{figure}

\subsection{Magnetization reversal process in pinwheel BASI}
\label{APP:config1} 
\begin{figure}[H]
\centering
\includegraphics[width=1.0\linewidth]{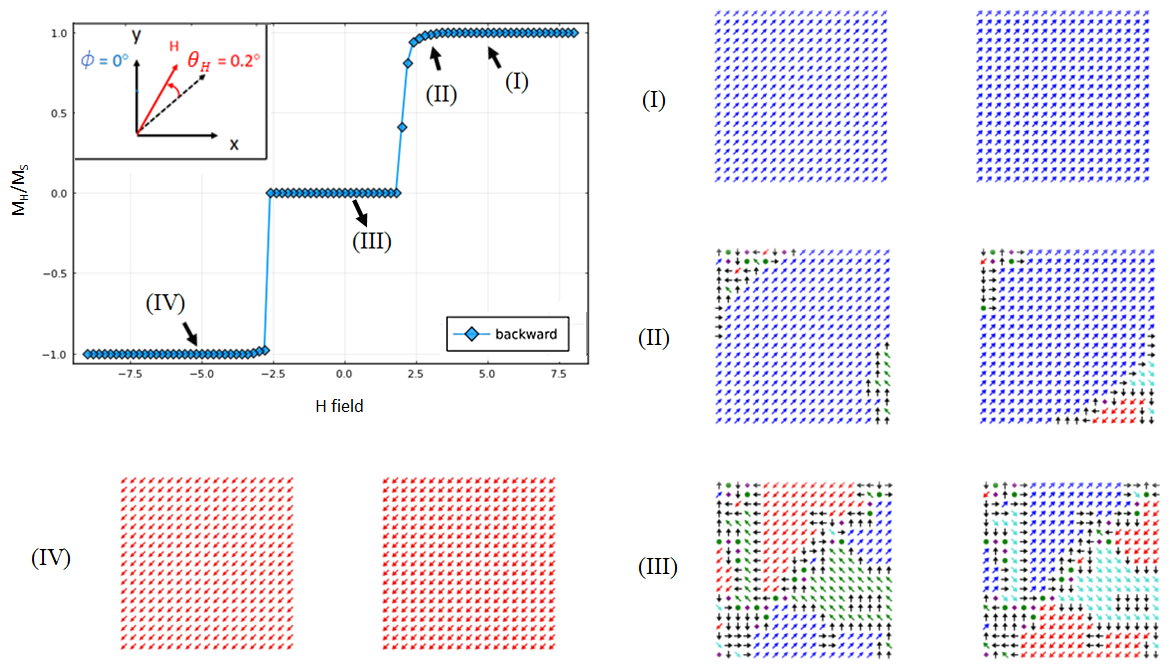}
\caption{Magnetization reversal process in a $\phi=0^{\circ}$ pinwheel BASI configuration subject to a field applied at an angle $0.2^{\circ}$ to the $xy$ diagonal. Four points are marked in the hysteresis loop across the reversal. The snapshots of the vertex configurations at each of these marked points are also given. Colour-coded arrows represent type T2 and T3 vertex magnetization orientations. The circles represent T1 vertices while T4 vertices are indicated by the diamond symbols.}
\label{fig:fig44}
\end{figure}

\begin{figure}[h!]
\centering
\includegraphics[width=1.0\linewidth]{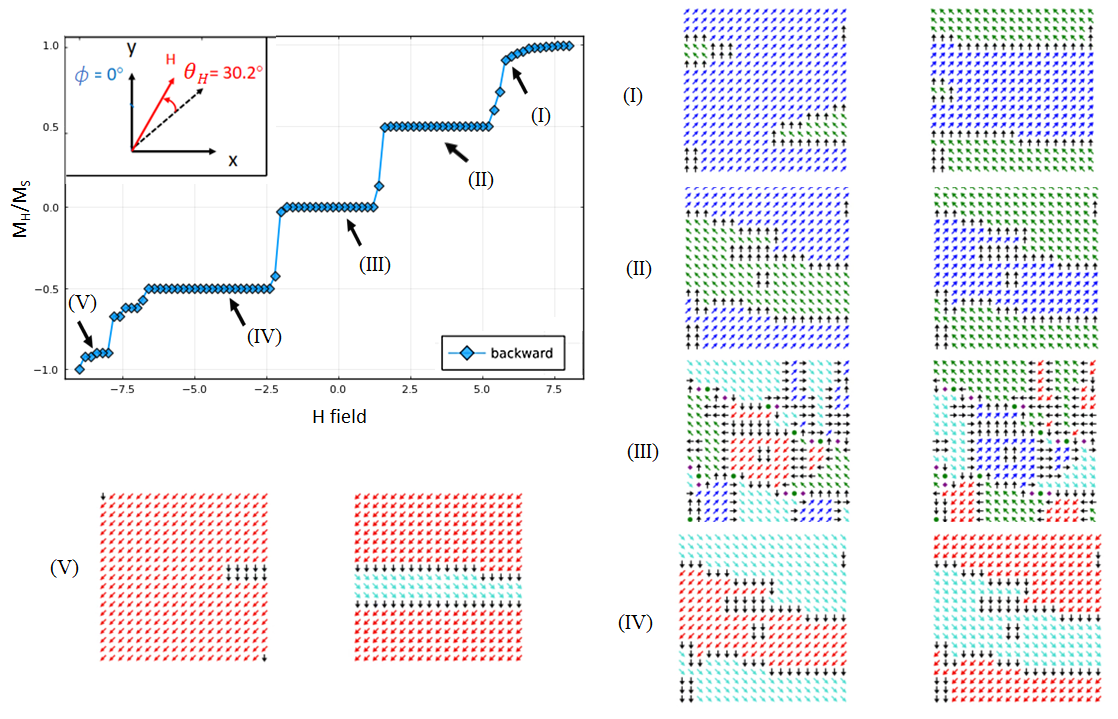}
\caption{Magnetization reversal process in a $\phi=0^{\circ}$ pinwheel BASI configuration subject to a field applied at an angle $30.2^{\circ}$ to the $xy$ diagonal. Five points are marked in the hysteresis loop across the reversal. The snapshots of the vertex configurations at each of these marked points are also given. Colour-coded arrows represent type T2 and T3 vertex magnetization orientations. The circles represent T1 vertices while T4 vertices are indicated by the diamond symbols.}
\label{fig:fig45}
\end{figure}

\begin{figure}[H]
\centering
\includegraphics[width=1.0\linewidth]{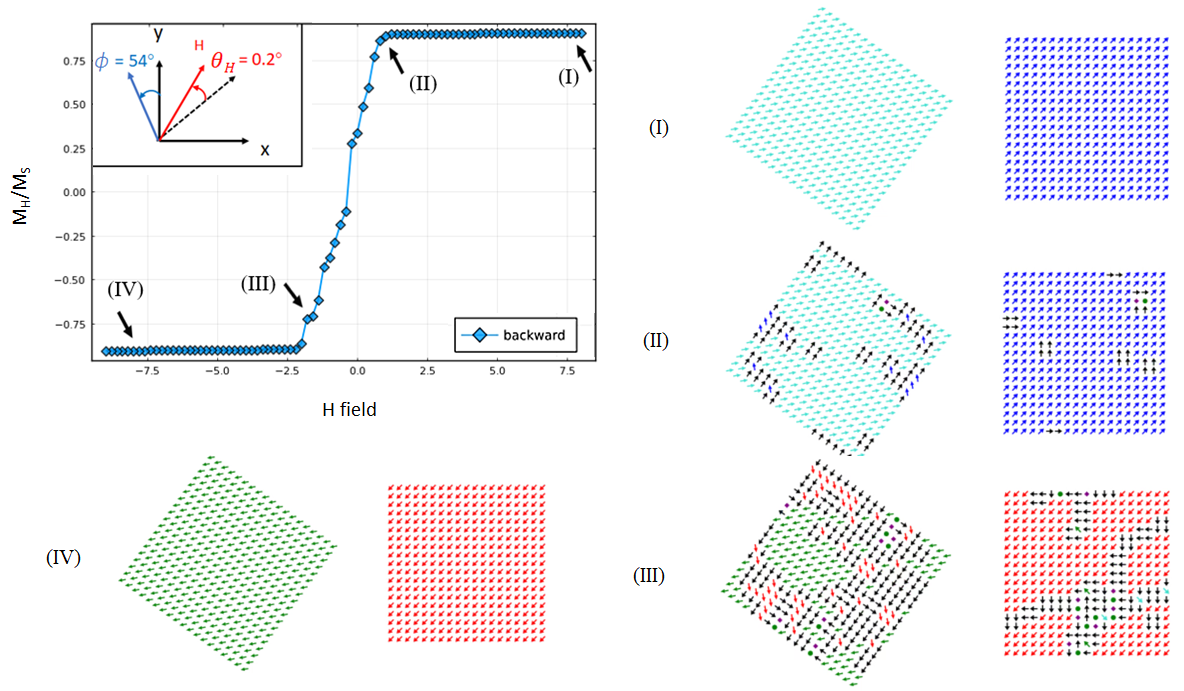}
\caption{Magnetization reversal process in a $\phi=54^{\circ}$ pinwheel BASI configuration subject to a field applied at an angle $0.2^{\circ}$ to the $xy$ diagonal. Four points are marked in the hysteresis loop across the reversal. The snapshots of the vertex configurations at each of these marked points are also given. Colour-coded arrows represent type T2 and T3 vertex magnetization orientations. The circles represent T1 vertices while T4 vertices are indicated by the diamond symbols.}
\label{fig:fig46}
\end{figure}

\end{document}